# A large-scale bibliometric analysis of global climate change research between 2001 and 2018

Hui-Zhen Fu[1*], Ludo Waltman[2]

[1] Department of Information Resources Management, School of Public Affairs, Zhejiang University, No. 866 Yuhangtang Road, Hangzhou 310058, China

[2] Centre for Science and Technology Studies, Leiden University, Leiden, The Netherlands; https://orcid.org/0000-0001-8249-1752

*Author to whom correspondence should be addressed

E-mail: fuhuizhen@zju.edu.cn

## Abstract

Global climate change is attracting widespread scientific, political, and public attention owing to the involvement of international initiatives such as the Paris Agreement and the Intergovernmental Panel on Climate Change. We present a large-scale bibliometric analysis based on approximately 120,000 climate change publications between 2001 and 2018 to examine how climate change is studied in scientific research. Our analysis provides an overview of scientific knowledge, shifts of research hotspots, global geographical distribution of research, and focus of individual countries. In our analysis, we identify five key fields in climate change research: physical sciences, paleoclimatology, climate-change ecology, climate technology, and climate policy. We draw the following key conclusions: (1) Over the investigated time period, the focus of climate change research has shifted from understanding the climate system toward climate technologies and policies, such as efficient energy use and legislation. (2) There is an imbalance in scientific production between developed and developing countries. (3) Geography, national demands, and national strategies have been important drivers that influence the research interests and concerns of researchers in different countries. Our study can be used by researchers and policy makers to reflect on the directions in which climate change research is developing and discuss priorities for future research.

## 1 Introduction

Climate change, also known as global warming, refers to an increase in the average global temperature, which can have severe effects on the planet (United Nations, 2020; TakePart, 2020). Since the pre-industrial era, there has been a significant increase in the average global temperature of 0.9 °C, which does not correspond to the natural cycles of the planet (NASA, 2020). The United Nations has highlighted climate change as one of the present day's major global issues, as the impacts of climate change are global in scope and unprecedented in scale. Some of the major consequences of climate change include shifting weather patterns that threaten food production, warming ocean temperatures that are associated with stronger and more frequent storms, rising sea levels that increase the risk of catastrophic flooding, heat waves that contribute to human deaths and other consequences, and increasing incidence and severity of wildfires that threaten habitats, homes, and lives (United Nations, 2020; TakePart, 2020). Although a small minority of researchers have questioned the reality of climate change, more than 97% of the scientific community affirm that climate change is occurring (Cook et al., 2013). Based on the 222,060 papers published between 1980 and 2014, it was reported that the total number of climate change related papers doubled every 5–6 years (Haunschild et al., 2016).

Obtaining a full overview of climate change research is important for both intellectual and political reasons. Scientific publications provide an important perspective to understand how the scientific community acts to tackle climate change, as publications reflect the priorities set by governments that fund climate change research and the research topics that scientists choose to focus on. However, the rapidly increasing number of climate change publications makes it challenging for researchers in this field to maintain an up-to-date overview of the literature. Bibliometrics is a powerful method for analyzing the development of scientific literature in a research field from a quantitative perspective, and it has been widely used in many global studies (Narin et al. 1976;

Callon et al., 1991; Coulter et al., 1998; Garcia-Rio et al., 2001; Vergidis et al., 2005; Fu et al., 2013; Li et al., 2020a). Owing to increasing public and scientific attention to climate change, previous researchers have successfully employed bibliometric methods to characterize the intellectual landscape of climate change. Nonetheless, most of these studies only focused on a specific research area within the field of climate research, such as the abstracting journal of the American Meteorological Society (Stanhill, 2001), climate change vulnerability (Wang et al., 2014), climate change controversy between the Intergovernmental Panel on Climate Change (IPCC) and contrarian reports (Janko et al., 2017), climate change and tourism (Fang et al., 2018), climate change adaptation (Wang et al., 2018), climate change research in the Arab world (Zyoud and Fuchs-Hanusch, 2020), climate change and carbon sinks (Huang et al., 2020), and climate change and infectious diseases (Li et al., 2020a).

Several previous bibliometric studies have presented comprehensive analyses of climate change research. In 1997, highly dynamic, rapidly developing fronts of climate research were identified based on 2,465 relevant publications in the Science Citation Index by using co-citation analysis (Schwechheimer and Winterhager, 1999). Global climate change research published between 1992 and 2009 has been assessed from the perspectives of research patterns, tendencies, and methods based on the Science Citation Index Expanded (Li et al., 2011). For climate change literature published between 1980 and 2015, the overall publication output, major subfields, and contributing journals and countries, as well as their citation impact, and a title word analysis have also been presented by using Web of Science (Haunschild et al., 2016). A bibliometric approach has been used to characterize climate change and global warming literature from 2001 to 2016 from the Web of Science database using growth trends, authorship patterns, and collaboration perspectives (Sangam and Savitha, 2019). However, as climate change research has developed rapidly over the past few years, it is necessary and interesting to use a more recent dataset to trace research evolution and research fronts. This will hopefully provide a better overview of the entire climate change research field as it currently stands.

There has been an increasing demand from researchers and policymakers to obtain an overview of developments in the field of climate change research. Terms and phrases in a publication's title, abstract, or author keywords have often been considered as a useful summary of the contents of scientific publications, especially for topic identification of research. Therefore, one technique developed to provide an overview of research fields is term mapping analysis, which counts and analyzes the co-occurrence of terms in scientific publications on a given subject (Callon et al., 1983). Term mapping analysis is based on the nature of terms, which are important carriers of scientific concepts, ideas, and knowledge (van Raan & Tijssen, 1993). Term mapping analysis has been widely used as a method to present an overview of the organization of scientific literature on a specific research topic, such as polymer chemistry (Callon et al., 1991), software engineering (Coulter et al., 1998), economics (Cahlik, 2000), robot technology (Lee and Jeong, 2008), life cycle assessment (Hou et al., 2015), environmental crisis management (Dai et al., 2020), sustainable design (Geng et al., 2020), and sustainable banking (Najera-Sanchez, 2020). Another important tool for bibliometric studies are burst detection algorithms, which are used to explore the development of a research field over time. For example, the burst detection algorithm proposed by Kleinberg (2003) has been used to detect the sudden increase in the frequency of use of subject terms and to identify hot topics or emerging trends (Mane and Börner, 2004; Swar and Khan, 2014; Marrone, 2020).

Increasing computing power combined with improved access to bibliographic data has fostered the growing popularity of bibliometric methods. Faster and more sophisticated algorithms have been developed and embedded in different tools, such as VOSviewer and the Science of Science tool (Sci2). VOSviewer offers text mining functionality to construct and visualize a term map (van Eck and Waltman, 2011), and Sci2 supports the use of burst detection algorithms (Sci2 Team, 2009).

These tools have enabled researchers from various disciplines to work with sophisticated algorithms for language processing, network analysis, and visualization.

In this paper, we aim to provide an overall picture of the climate change field from a bibliometric perspective, contributing to a fuller and deeper understanding of the main features and publication patterns of the field. In particular, we focus on (1) the overall structure of climate change research and the main research topics by using term mapping analysis, (2) the historical development, research shifts, and emerging research topics in climate change research by using burst detection analysis, and (3) national attention to climate change and national concerns.

## 2 Methodology

### 2.1 Search strategy

Our bibliometric analysis focused on publications from the period 2001–2018. Web of Science (WoS) and Scopus are the two most common literature databases for bibliometric studies and can usually be expected to yield fairly similar results in large-scale analyses (Torres-Salinas et al., 2009; Archambault et al., 2009). They have both been used in previous bibliometric studies on climate change because of their broad coverage of various disciplines (Li et al., 2011; Wang et al., 2014). WoS contains more than 12,000 academic journals, and essentially covers the international authoritative academic journals that publish climate change research. Our analysis was based on the in-house WoS database available at the Centre for Science and Technology Studies (CWTS) at Leiden University. This database covers the Science Citation Index Expanded, the Social Sciences Citation Index, and the Arts & Humanities Citation Index.

To identify publications dealing with climate change, we developed a keyword-based search strategy. This strategy was based on a literature review and comprehensive consideration of search strategies from previous bibliometric studies on climate change (Li et al., 2011; Wang et al., 2014; Zyoud and Fuchs-Hanusch, 2020). We searched for the keywords “climate chang*,” “climatic chang*,” “climate variabilit*,” “climatic variabilit*,” “global warming,” “climate warming,” and “climatic warming” in the titles and abstracts of publications in WoS from 2001 to 2018. Only publications classified as articles were considered. Other types of publications, such as reviews, notes, and conference papers, were excluded. Our search strategy identified 122,815 publications on climate change.

### 2.2 Term mapping analysis

Term mapping analysis focuses on analyzing the number of documents in which pairs of terms appear together, thus providing insights into the knowledge structure and research frontier of a certain subject. The results of a term mapping analysis are usually presented in a network visualization, which is a powerful tool for exploring the structure and dynamics of scientific fields (Börner, 2010). In this study, we used the VOSviewer software for bibliometric visualization, developed at CWTS, Leiden University (www.vosviewer.com; van Eck and Waltman, 2010). Term co-occurrence analysis, construction of a term map, and visualization of the map was carried out in the following five steps.

Step 1: A random sample of climate change publications was selected to reduce the computational requirements as well as the memory requirements of the term co-occurrence analysis. We randomly selected 25,000 articles from the entire dataset of 120,815 articles on climate change research. A sample of 25,000 publications is sufficient to obtain robust results.

Step 2: The most characteristic terms of climate change research were identified using a natural language processing approach. A list was then created of all noun phrases that occurred in the titles and abstracts of the 25,000 climate change publications selected in Step 1. A subset of the more frequently occurring noun phrases was selected as the most characteristic terms of climate change research (van Eck and Waltman, 2011), leaving out general noun phrases such as “result,” “study,” and “patient.” Noun phrases occurring in fewer than 10 articles were excluded because they are likely to be of little interest and may complicate the analysis. For each of the remaining noun phrases, VOSviewer calculated a relevance score. Based on these scores, we selected the 60% most relevant noun phrases, an approach that typically works reasonably well (van Eck & Waltman, 2011). This resulted in a final set of 7,060 noun phrases that could be regarded as important and relevant terms in the field of climate change.

Step 3: The network of co-occurrences of terms was constructed from the 7,060 terms selected in Step 2, by determining the number of publications each pair of terms appeared in. Two terms were deemed to co-occur in a publication if they both occurred at least once in the title or abstract of the publication. The number of co-occurrences of two terms can be interpreted as a measure of the relatedness of the terms.

Step 4: A term map was created using VOSviewer's layout technique by positioning the terms in such a way that in general strongly related terms were closer to each other, while terms that did not have a strong relation were farther from each other. We used the clustering technique in VOSviewer to partition the terms in the term map into several clusters. In general, terms assigned to the same cluster were more strongly related, whereas terms assigned to different clusters were less strongly related.

Step 5: The term map was visualized using two types of visualizations: a density visualization and an overlay visualization. A density visualization, which uses colors to highlight the most prominent areas of interest in a term map (van Eck and Waltman, 2014), was used to reveal how the focus of climate change research has changed over time. An overlay visualization, which uses colors to indicate a certain property of the terms in a term map (van Eck and Waltman, 2014), was used to analyze the differences in the research focus of different countries.

### 2.3 Burst detection analysis

While term mapping analysis reveals how terms relate to each other, burst detection analysis identifies terms that show significant growth during a certain period. We used burst detection analysis to identify topics that had seen sudden growth, in order to highlight the latest trends in climate change research. To perform burst detection analysis, we used Sci2 (Sci2 Team, 2009), which includes Kleinberg's burst detection algorithm (Kleinberg, 2003). The temporal bar graph was used to show the temporal distribution and the time span of burst author keywords. The parameters were set as follows: Gamma = 1.0, Density Scaling = 1.2, Bursting States = 1, and Burst Length = 1 year.

# 3 Results and Discussion

## 3.1 Main climate change research topics by term map

Five clusters of research topics were identified based on the co-occurrence network of terms in climate change publications, as shown in Figure 1. The most frequently used term, "panel," refers primarily to the IPCC. The IPCC played an important role in combining the understanding of the physical sciences behind climate change (blue cluster) with the understanding of climate policy (green cluster) and climate technology (purple cluster) via its Fourth Assessment Report in 2007 and its Fifth Assessment Report in 2014. The importance of the IPCC in climate change research was not surprising as its purpose is to provide governments at all levels with scientific information to develop climate policies (IPCC, 2020). The IPCC is a group of more than one thousand scientists independent from governments or companies worldwide, and its reports are a key input for international climate change negotiations.

### *3.1.1 Physical sciences*

The dark blue cluster, located on the bottom-left side of the term map (Fig. 1), covers the physical sciences that are the foundation of climate change research. A key part of this field is the climate system, which involves Earth–Sun interactions as well as the interactions of matter, energy, and processes within the Earth's atmosphere, hydrosphere, cryosphere, lithosphere, and biosphere (Barry and Hall-McKim, 2014).

Many modeling-related terms can be found among the most frequently used terms. These include, "general circulation model (GCM)," "global climate model," "regional climate model (RCM)," "hydrological model," "atmospheric circulation," "precipitation change," "earth system model,"

“climate simulation,” and “climate model simulation.” This shows that numerical models, representing physical processes in the atmosphere, ocean, etc., are important tools for simulating the response of the global climate system to increasing greenhouse gas concentrations. Oceans occupy 70% of the Earth’s surface and are therefore of critical importance to the climate system. The climate system also includes terms with a large spatial scale, for example “Northern Hemisphere,” “Southern Hemisphere,” oceans such as the “North Atlantic,” “Indian Ocean,” “Pacific Ocean,” “Atlantic,” “North Atlantic,” and “Pacific,” and finally, the region of “East Asia.” Ocean-atmosphere interaction effects are also abundant terms; for example, the spontaneous generation of the El Niño/Southern Oscillation (ENSO). Terms such as “El Niño Southern Oscillation (ENSO)” and “El Niño” represent the most dominant feature of cyclic climate variability on sub-decadal timescales (Yeh et al., 2009). Because of the complexities of the physical processes involved in El Niño, researchers have heavily relied on complex climate models that represent interactions between these processes (Trenberth and Hoar, 1997).

*3.1.2 Paleoclimatology*

Paleoclimatology (yellow cluster, Fig. 1) occupies the middle-left side of the term map. This research field aims to understand changes in the climate system from a chronological perspective, as indicated by frequently occurring terms such as “record,” “sequence,” “archive,” and “chronology,” which are obtained from long-term records identified by terms such as “lake,” “sediment,” “deposition,” “glacier,” “sediment core,” “vegetation change,” and “pollen.” Paleoclimatology covers different periods of the Earth’s history, as shown by terms such as “ice age,” “Holocene,” “late Holocene,” “Pleistocene,” “early Holocene,” and “Last Glacial Maximum.” Studies of past changes in the climate system could help understand the causes of abrupt climate change and provide insights into the present-day climate.

Another frequently used term is “Tibetan Plateau,” which as the “roof of the world” exerts a strong influence on regional and global climate through thermal and mechanical forcing mechanisms (Kang et al., 2010). Many studies have focused on the impact of climate change on the Tibetan Plateau's energy and water cycle (Yang et al., 2014).

*3.1.3 Climate-change ecology*

Climate-change ecology (red cluster, Fig. 1) is the largest field and occupies the top-left side of the term map. The climate has an important environmental influence on ecosystems, and changes to the climate can affect ecology in various ways. "Survival" is an important topic in this field, together with "population dynamics" and "treatment," because the living environment for animals and plants deteriorates as the climate changes. Climate change has a profound influence on "community structure" and "community composition" and subsequently has an impact on ecosystem functioning and feedback to climate change. Prominent terms in this field include "taxa," "tree species," "bird," "species distribution," "fauna," "species richness," "species composition," "insect," "plant species," "plant community," and "leaf."

The agricultural perspective is also an important aspect of climate-change ecology. Mitigation of greenhouse gas emissions and adaptation to climate change can increase farm productivity and sustainably. Agriculture is a major part of the climate problem and generates three primary greenhouse gases: $CO_2$, $CH_4$, and $N_2O$ (Johnson et al., 2007). It is also the main source of $N_2O$ emissions, accounting for approximately 75% of the world $N_2O$ emissions (Olivier and Peters, 2018). How agriculture can reduce its greenhouse gas emissions through conservation measures is an important research topic (Johnson et al., 2007), as reflected by terms such as "nitrogen," "methane," "$CH_4$," "organic matter," "$N_2O$," "organic carbon," "soil organic carbon," "nitrous oxide," "$N_2O$ emission," "$CH_4$ emission," "oxidation," "soil organic matter," "methane emission," "carbon flux," and "carbon balance." Climate change can have a major effect on the global food supply and could potentially cause famine (Parry et al., 2005). Therefore, to maintain and increase farm productivity and sustainability, "rice," "maize," "wheat," "crop yield," "crop production," "grain yield," "net primary productivity," and "crop model" are important concerns for researchers. Moreover, there has been increasing attention on some important grain-producing regions to obtain a better understanding of the impact of climate change and how to adapt to it. An example of such a region is "Northeast China," the main crop production region in China.

*3.1.4 Climate technology*

Climate technology (purple cluster, Fig. 1) occupies the top right side of the term map. This cluster focuses on the mitigation of climate change from the perspective of technology. As expected, there are several terms relating to the targets of climate technologies to reduce greenhouse gas emissions. Examples include "emission," "gases," "carbon dioxide," "greenhouse gas emission," "$CO_2$ emission," "greenhouse gas," "GHG," and "GHG emission." These terms are located near the center of the map.

Stabilizing greenhouse gas (GHG) concentrations at a safe level for the climate system was the objective of the United Nations Framework Convention on Climate Change (UNFCCC) in 1992 (United Nations, 1992), the predecessor of the Paris Agreement. A commonly used metric, "global warming potential" or "GWP," is usually used to transform the effects of different greenhouse gas emissions to a common scale (IPCC, 2014). Since $CO_2$ emissions from fossil fuel combustion and industrial processes contribute to approximately 78% of the total GHG emission increase (IPCC, 2014), current climate technologies usually focus on the more efficient use of energy and switching to low-carbon fuels, such as renewable energy. Relevant terms include "energy," "energy efficiency," "technology," "industry," "consumption," "energy consumption," "electricity," "fossil fuel," "fuel," "oil," "coal," and "alternative."

"Environmental impact" is another topic related to assessing the expected environmental impact of decisions or policies. More recently, "life cycle assessment" and "LCA" have been concerned with reviewing the environmental impact of products throughout their lives.

*3.1.5 Climate policy*

Climate policy (green cluster, Fig. 1) is the second largest cluster and occupies the bottom-right side of the term map. There is a strong scientific consensus that global warming is primarily caused by human activities. This consensus is demonstrated by a highly cited article that quantified the scientific consensus on global warming in nearly 12,000 scientific publications from 1991 to 2011 (Cook et al., 2013). Support for the consensus has increased over time, and it has been endorsed by approximately 97% of these 12,000 publications (Cook et al., 2013). Accordingly, terms related to mitigation and adaptation, such as “practice,” “action,” “solution,” “mitigation,” “sustainability,” “climate change mitigation,” and “adaptation strategy,” are common. A global review of national laws and policies on climate change adaptation reported that more than 170 countries have introduced national policies and laws on climate change adaptation to address the impact of climate change (Grantham Research Institute on Climate Change and the Environment, 2019). This report also pointed out that adaptation to the impact of climate change has typically received lesser attention than mitigation in the global policy discourse, and there has been increasing recognition of the urgent need for adaptation.

Given the importance of national policy and action for global climate efforts, it is not surprising that within the climate policy cluster, “country” is the most frequently occurring term. The implementation of mitigation and adaptation depends on political debate and decision making, reflected by terms such as “issue,” “policy,” “article,” “decision,” “planning,” “implementation,” “goal,” “concept,” and “action.” There is also a high dependency on different actors operating at different levels, indicated by terms such as “country,” “government,” “society,” “city,” “sector,” and “person.”

The success of implementing climate policies is linked to the way they are integrated with policies of different levels of government and with civil society (de Oliveira, 2009). However, economic activity is also one of the most important drivers of increase in $CO_2$ emissions from fossil fuel combustion and thus is an important concern in climate policies (IPCC, 2014). In the term map, frequently used terms such as "cost," "economy," "benefit," "market," and "investment" are located close to the climate technology cluster.

**3.2 Shift of climate change research focuses between 2001 and 2018**

Annual climate change publication output has increased from 1,582 articles in 2001 to 15,311 articles in 2018, while the percentage of climate change publications compared to the total number of scientific publications on WoS has grown from 0.20% to 0.95% (Fig. 2). This clearly shows how climate change has received increasing attention from the scientific community since 2001. It is common for new branches of science to continually evolve, while others gain or lose importance, merge, or split (Mane and Börner, 2004). The three term map density visualizations presented in Figure 3 show temporal shifts in the focus of climate change research.

We divided the entire investigation period of 2001–2018 into three equal spans of sub-periods: 2001–2006, 2007–2012, and 2013–2018. The terms and their locations are identical in the three visualizations in Figure 3; however, the number of occurrences of each term changed between the three time periods as shown by changes in color. Terms that occur frequently create regions with a stronger yellow color, whereas terms that occur less frequently create regions of green or blue. These three visualizations show the primary focus of climate change research in each time period and how the focus of climate change has shifted from understanding the climate system to adaptation and mitigation of climate change.

To explore new research trends in more detail, we used the burst detection algorithm embedded in the Sci2 tool to show the temporal distribution of burst author keywords. Considering the visualization effect, the top 30 keywords with the largest burst weights were selected to conduct burst detection analysis, as shown in Figure 4. Each horizontal bar represents a burst keyword, with the length of the bar representing the burst duration, the two sides representing a specific start and end year, and the area representing the burst weight of a keyword. The Kyoto Protocol, which entered into force in 2005, Copenhagen Accord (2009), and the Paris Agreement (2016) all appear to have played an important role in the development of climate change research between 2001 and 2018.

*3.2.1 Period 2001–2006*

The number of climate change publications increased from 1,582 in 2001 to 2,776 in 2006, an average growth of 239 publications per year. Paleoclimatology was the research field that gained the most attention between 2001 and 2006 (Fig. 3).

During this period, a large number of burst terms were produced (Fig. 4.), most of which were relevant to paleoclimatology (e.g., “Holocene,” “paleoclimate,” “diatoms,” “pollen,” “quaternary,” “palynology,” “Pleistocene,” “stable isotopes,” and “phylogeography,”). This is consistent with the results shown in Figure 3 and confirms that research during this period focused mainly on paleoclimatology. General terms such as “greenhouse gases,” “carbon dioxide,” “climate variability,” “human impact,” and “modeling” have been attracting attention for a long time.

In 2005, the Kyoto Protocol operationalized the UNFCCC by getting industrialized countries to commit to limit and reduce greenhouse gas emissions in accordance with agreed individual targets (United Nations Framework Convention on Climate Change, 2020a). The protocol has brought widespread discussion and emerging attention to climate change research before and after its enforcement and has become a hotspot in the field of climate change studies between 2001 and 2011. In addition, 2001–2006 was the period in which the terms “el Niño,” “fire,” “Canada,” “North Atlantic Oscillation,” and “extinction” emerged, indicating that interest in climate change research and the diversity of topics within the field was growing.

*3.2.2 Period 2007–2012*

Climate change studies began to develop rapidly during this period, with the number of climate change publications growing from 3,565 in 2007 to 8,801 in 2012. This is an average growth of 1,004 publications per year—four times higher than that in the period 2001–2006.

In addition, 2007–2012 is the period in which the Copenhagen Accord was signed (2009). The Accord was a voluntary agreement between over 180 countries, representing almost 80% of global emissions (United Nations Framework Convention on Climate Change, 2009).

The period between 2007 and 2012 appears to have been a transition period in climate change research with research topics such as climate technology and climate policy gaining more attention than paleoclimatology and the physical sciences (Fig. 3).

The percentage of anthropogenic global warming endorsements increased marginally from 1991 to 2011, approaching approximately 98% in 2011 (Cook et al., 2013). Nonetheless, there was a discrepancy between public opinion on climate change and the view of climate scientists; for example, only approximately half of the American public believed in anthropogenic climate change at the time (Leiserowitz et al., 2013). However, public awareness of climate change in the United States was increased by Al Gore and the IPCC, who cited climate change as a threat to international security. For this work, they were awarded the Nobel Peace Prize in 2007.

Most burst terms during the 2001–2006 period were ongoing during the 2007–2012 period, while only a small number of new terms, such as “biodiversity,” “adaptation,” and “food security,” appeared (Fig. 4). Based on an analysis of the publications of researchers worldwide, food security in the field of agricultural systems and water resource management have been considered the most important research topics in climate change vulnerability (Wang et al., 2014). Our results indicate that the focus of climate change research has gradually shifted from understanding the climate system to adaptation of ecosystems and human beings to climate change.

*3.2.3 Period 2013–2018*

The number of climate change publications grew from 10,006 in 2013 to 15,311 in 2018, an average growth of 1,085 publications per year, similar to the growth seen between 2007 and 2012. Research topics such as climate technology and climate policy grew very quickly and became prominent contributors to climate change research (Fig. 3). We note that while paleoclimatology was not the most studied field in climate change research between 2013 and 2018, the number of yearly paleoclimatology publications increased consistently during our period of analysis (2001–2018), but with a lower growth rate than the topics of climate technology and climate policy.

The burst weights of terms were also greater between 2013 and 2018 than they were between 2001 and 2012. Between 2013 and 2018, there was a scientific consensus that human-caused climate change was occurring (Maibach et al., 2014), and new terms with high burst weights appeared in climate change research, indicating the emergence of new research branches. Among them, the burst weight of “climate change adaptation” and “climate change mitigation” was 20.2 and 9.1, respectively, which reflects the sharp increase in research into adaptation and mitigation of climate change during this period. These two burst terms are related to two broad categories of policy solutions to address climate change issues. The first was a set of strategies to adapt to the effects of climate change, with the goal of reducing the damage caused in the present and future. The second was a set of interventions designed to mitigate the production of greenhouse gases with the goal of slowing and eventually halting the advancement of climate change.

The Paris Agreement in 2016 saw nations agree to undertake ambitious efforts to combat climate change and adapt to its effects for the first time (United Nations Framework Convention on Climate Change, 2020b) and charted a new course in the global climate effort. The term “Paris Agreement” first appeared in the literature in 2017.

Modeling climate is an important part of climate change research. The World Climate Research Programme Working Group on Coupled Modelling agreed to promote a new set of coordinated climate model experiments in 2008, comprising the fifth phase of the Coupled Model Intercomparison Project (CMIP5) (Taylor et al., 2012). The burst weight of “CMIP5” was 15.2 in the period 2013–2018, which shows that CMIP5 has had a large scientific impact in recent years.

Research related to ecosystem services has grown quickly from 2013 to 2018 with a burst weight of 7.9. The sustainability of ecosystem services has received an increasing amount of scientific attention to better understand how organisms provide these services to humans and how these organisms are affected by a changing climate (Prather et al., 2013). “Life cycle assessment,” which usually relates to projects for human beings, had a high burst weight of 17.9 after 2016, consistent with the research shifts visible in Figure 3.

The term “resilience” has been increasingly employed, with a burst weight of 13.5 after 2014. The term is used not only in the context of the study of natural systems, such as coastal marine ecosystems (Bernhardt and Leslie., 2013), coral reefs (Mumby et al., 2014), and forests (Stevens-Rumann et al., 2018), but also in the study of man-made systems, such as cities (Satterthwaite, 2013; Boyd and Juhola, 2015). The term “resilience” is increasingly used in the context of discussions, policies, and programming around climate change adaptation and disaster risk reduction (Bahadur et al., 2013). The 10 key characteristics of resilience have been summarized as “high diversity; effective governance and institutions; the ability to work with uncertainty and change; community involvement and the inclusion of local knowledge; preparedness and planning for disturbances; high social and economic equity; robust social values and structures, acknowledging non-equilibrium dynamics, continual and effective learning, and the adoption of a cross-scalar perspective” (Bahadur et al., 2013).

## 3.3 Distribution of global scientific attention and research focus of top eight countries

### *3.3.1 Distribution of global scientific attention*

Climate change is a global issue that concerns all of the world’s nations and governments. The global geographical distribution of climate change research is shown in Figure 5. More than 200 countries/territories worldwide published climate change research between 2001 and 2018, indicating a global concern from the scientific community. Eight countries published more than 10,000 articles: the USA, China, the UK, Australia, Germany, Canada, France, and Spain. The USA was the most active in global climate change research, with a contribution from the USA to 73% (89,260 articles) of the total published articles, followed distantly by China (29%; 35,834 articles) and the UK (23%; 28,166 articles). The USA published the largest number of climate change articles each year between 2001 and 2018, while China's number of publications increased rapidly since 2010 and surpassed the number of publications by the UK in 2014 (Fig. 6).

A study in 2015 found an increase in climate change awareness among adults in developed countries based on the results of a Gallup World Poll in 2007–2008 that covered 119 countries (Lee et al., 2015). It was shown that across much of North America and Europe, over 90% of respondents were aware of climate change. In contrast, a much lower percentage of respondents were aware of climate change in developing countries—35% of respondents in India and 25% of respondents in Egypt (Lee et al., 2015). Another cross-national social survey also revealed that national wealth has a positive correlation with the perceived importance of climate change and the awareness of the danger of climate change (Lo and Chow, 2015). To some extent, global public attention was roughly in line with scientific attention regarding climate change issues.

*3.3.2 Research focus of top eight countries*

The term map overlay visualizations presented in Figure 7 show the research focus of the eight countries with the largest number of climate change publications. The color of a term indicates the share of publications with the term that a country has produced. Terms with a low publication share are white or light blue, while terms with a high publication share are dark purple. For example, in the case of the USA, "el Niño southern oscillation" is dark purple, which means that at least half of the global publications in which "el Niño southern oscillation" occurs were produced by researchers in the USA. It should be noted that owing to the large differences in the number of climate change publications produced by different countries, the different overlay visualizations in Figure 7 use different color scales.

The visualizations provide a picture of each country's research focus at the macro level, but a country's most important terms are not easy to identify. To explore the research focus of the different countries in detail, the top 200 terms that occurred in more than 203 publications were compared for each country. Of these terms, the 10 that received the highest publication share for each country are listed in Table 1. Only the top 200 terms were selected to ensure that only important terms were included in the comparison. The USA had the highest publication share for most terms in Table 1 as it produced the largest volume of publications. Overall, the USA was slightly more focused on the physical sciences and climate-change ecology than on paleoclimatology, climate technology, and climate policy.

Geography appears to be an important consideration for most countries, as the top term with the highest publication share for six countries had a clear geographical focus; “El Niño southern oscillation” for the USA (54%), “Tibetan plateau” for China (89.9%), “North Atlantic” for the UK (28.4%) and Germany (21.8%), “permafrost” for Canada (22.4%), and “Atlantic” for France (16.4%). Most of these geographical terms were relevant to the physical sciences, paleoclimatology, and climate-change ecology, which is understandable owing to the nature of these disciplines. It is noticeable that China is responsible for the vast majority of research on the Tibetan Plateau (89.9%). The Tibetan Plateau, the vast high plateau of southwestern China, the ‘third pole’ of the Earth, and the world’s largest plateau, has experienced approximately double the observed global average rate of global warming, resulting in significant permafrost thawing and glacier retreat (Chen et al., 2013). A review by Chen et al. (2013) summarized several suggestions to reduce the uncertainties and improve the precision of predictions of the impacts of climate change and human activities on the biogeochemical cycle of the Tibetan Plateau.

There are three classic geographic perspectives: (i) looking at the world through the lenses of place, space, and scale; (ii) different domains of dynamics, including human-societal, environmental, and environmental-societal dynamics; and (iii) spatial representation using visual, verbal, mathematical, digital, and cognitive approaches (National Research Council, 1997). Geographic perspectives in climate change research have long been understood to be of spatial, historical, and cultural importance across the natural sciences, social sciences, and humanities (Ackerly et al., 2010; Brace and Geoghegan, 2011; Yang et al., 2014). However, researchers are often more likely to act from the perspective of national security or national demands.

China's focus on climate change research was more varied than other countries, as shown in Figure 7, where the deep purple region occupies the upper middle of the term map. Although China has focused heavily on climate-change ecology, this is more from an agricultural perspective than a natural perspective. After "Tibetan Plateau," China's highest publication shares include terms such as "rice," "soil temperature," "$N_2O$," "carbon emission," and "utilization," which is consistent with China's research focus on agriculture (Fig. 7). Agriculture in the context of global warming includes two perspectives: mitigation by reducing greenhouse gas emissions and adaptation to climate change to ensure food supply. Since China accounts for 20% of the world's population, food security has long been a major concern in its national strategy. The State Council of China has claimed that "as food decides national prosperity and the people's wellbeing, food security is a major prerequisite for national security" (2019). In 2008, China introduced the first national mid- to long-term food security plan (2008–2020), in which the government reiterated its commitment to achieving a 100% self-sufficiency rate in the supply of cereals (State Council, 2008). Rice is ranked first among cereal crops in China, and between 2019 and 2020, China consumed significantly more rice than any other country (Statista, 2020). However, global warming and natural disasters have made it more difficult for China to ensure food security. Therefore, exploring the impact of climate change on the rice production potential in China is of great significance in adapting to climate change and safeguarding China's food security (Yao et al., 2007; Chen et al., 2020; Li et al., 2020b). Such research could also benefit other countries and help safeguard food security on a global scale. To achieve this, priority should be given to adjusting the structure of agricultural cultivation and appropriately expanding the planting area of maize and rice to adapt to climate change, thus accelerating the construction of agricultural infrastructure to reduce the negative impact of declining precipitation and increasing daytime temperature on agriculture (Li et al., 2020b).

National strategies, such as national climate laws and policies, have also influenced the choice of research topics by climate change researchers. The UK and Australia have an obvious focus on climate policy, with eight of the top 10 terms from the UK being “actor,” “priority,” “governance,” “carbon emission,” “initiative,” “intervention,” “target,” and “society,” and seven of the top 10 terms from Australia being “climate change adaptation,” “adaptive capacity,” “health,” “person,” “government,” “livelihood,” and “benefit.” Moreover, “actor” related publications from the UK represented the highest publication share worldwide. According to the Climate Change Laws of the World database (2020), the European Union has the largest number of climate laws and policies in the world. The EU member states reported that they had already adopted or were planning to adopt 1,925 national climate policies and corresponding measures in 2019, many of which will help achieve energy efficiency and renewable energy targets (European Environment Agency, 2019). The strong contribution of UK researchers is not surprising because the UK has been recognized as a global leader in climate change with ambitious and legally binding emission reduction targets, supported by highly developed policies, legislation, and institutions (Sustainable Prosperity, 2012). The UK is the most active European country in terms of implementation of climate change policies, with 80% of its cities having an adaptation strategy, as shown by a comparative study of climate change strategies and plans (Heidrich et al., 2016). The landmark piece of UK legislation was the 2008 Climate Change Act, which represents the world’s first legally binding national commitment to cut greenhouse gas emissions. It commits the government to cut national greenhouse gas emissions by at least 100% of the levels in 1990 (net zero) by 2050 (Energy & Climate Intelligence Unit, 2020). The political and institutional circumstances surrounding the UK climate politics and the Climate Change Act have received increasing attention (Carter, 2014; Lorenzoni and Benson, 2014; Gillard, 2016). Canada summarized the lessons from the UK as eight elements: long-term low-carbon transition plans, suites of policies, short-term targets, legal frameworks, coordinating bodies/departments, independent advisory bodies, green investment banks, and economic reviews of climate change (Sustainable Prosperity, 2012).

According to the Climate Change Laws of the World database (2020), Australia has the most litigation cases (107) globally, followed by the UK (64). This indicates that the Australian Government had a strong determination to address the issue of climate change. Australia came very close to legislating an emissions trading scheme as part of a climate policy package in 2009, driven by Prime Minister Kevin Rudd, whose climate policy was underpinned by ecological modernization and climate justice (Curran, 2011). The Australian Government released a National Climate Resilience and Adaptation Strategy in 2015, which identified a set of principles to guide effective adaptation practices and resilience building and outlined the Australian Government's vision for a climate-resilient future.

## 4 Conclusions

Our bibliometric analysis of climate change research in the period 2001-2018 has used term maps to reveal five main research fields and their evolution on a global scale. We have also determined how research into these fields varies worldwide and how different countries focus on different areas of climate change research. Our analysis may serve as a reference for readers, researchers, and policymakers interested in understanding the areas of climate change research that are the most relevant.

Based on the clusters of terms in our term map, we have identified the following main fields of climate change research: physical sciences, paleoclimatology, climate-change ecology, climate technology, and climate policy. The scientific community has reached a scientific consensus on anthropogenic global warming (Cook et al., 2013), and as we have shown, between 2001 and 2018 the scientific community has strongly increased the amount of attention that it gives to climate change. International agreements and initiatives such as the Kyoto Protocol, the Copenhagen Accord, the Paris Agreement, and the IPCC have played an important role in addressing climate change.

Through density visualizations of term maps, we have identified changing trends in the main topics in climate change research. Over the investigated period, the focus of climate change research has shifted from understanding climate systems, such as paleoclimate, toward climate change adaptation and mitigation, including climate technologies and climate policies, such as using energy more efficiently, switching to renewable energies, and encouraging the development of national policies and laws.

Climate change is a global issue and has received global scientific attention, but the ability to tackle climate change issues is not evenly distributed worldwide. There was an imbalance in scientific production between developed and developing countries, with developed countries contributing much more than developing countries. The USA contributed the most to climate change research, accounting for approximately 73% of all climate change publications.

A diverse set of perspectives such as geography, national demands, and national strategy have been found to drive scientists from different countries to devote themselves to specific topics in climate change research. (1) Geography is an important perspective for researchers from different countries, and it shapes national climate change research with a focus on local issues, such as the El Niño Southern Oscillation for the USA, the Tibetan Plateau for China, the North Atlantic for the UK and Germany, and permafrost for Canada. (2) National demands call for scientists' attention on a large scale at the national level. For example, Chinese researchers have focused on rice and agriculture, because food security is a major prerequisite for national security in China and because priorities have been given to agricultural cultivation and appropriately expanding the planting area of maize and rice to safeguard food security. (3) National strategies, including national climate laws and policies, suggest that the research direction for several countries lean towards strategic and long-term perspectives for climate change researchers. For example, the UK contributed much more to climate policy research, which could be partly because of the UK's 2008 Climate Change Act, the world's first legally binding national commitment to cut greenhouse gas emissions.

It should be noted that our study has several limitations. (1) Although WoS is a commonly used literature database with a broad coverage of various disciplines, many journals, often with a local or regional focus, are not indexed in this database and are therefore not considered in our study. (2) Our study does not cover the most recent research on climate change. (3) Our analysis offers a macro level perspective on climate change research. It does not provide any detailed insights into sub-topics in climate change research at the micro level. More detailed insights would require combining a bibliometric analysis with an in-depth literature review.

## Acknowledgment

We would like to thank Helene Muri, Anders Hammer Strømman, and David Stern for their valuable comments on some of the results presented in this paper.

## Funding information

This study was supported by the National Natural Science Foundation of China (No. 71804163).

## Data availability

Our research is based on data from the Web of Science database. This is a proprietary database. We are not allowed to share the data used in our research.

**Table 1 Top 10 terms (among the global top 200 terms) with the highest publication share for each of the top eight countries**

| **USA** | **Topic** | **P** | **%** | **Max %** | **China** | **Topic** | **P** | **%** | **Max %** |
|---|---|---|---|---|---|---|---|---|---|
| el Niño southern oscillation | Physical science foundation | 304 | 54.0 | **54.0** | Tibetan plateau | Physical science foundation | 217 | 89.9 | **89.9** |
| pacific | Physical science foundation | 220 | 52.3 | **52.3** | rice | Climate technology | 204 | 34.3 | **34.3** |
| ENSO | Physical science foundation | 346 | 50.6 | **50.6** | soil temperature | Climate-change ecology | 244 | 34.0 | **34.0** |
| predator | Climate-change ecology | 206 | 50.0 | **50.0** | soil organic carbon | Climate-change ecology | 207 | 30.9 | 32.4 |
| scientist | Climate policy | 262 | 50.0 | **50.0** | $N_2O$ | Climate technology | 293 | 28.3 | **28.3** |
| streamflow | Physical science foundation | 253 | 48.2 | **48.2** | carbon emission | Climate policy | 279 | 27.2 | **27.2** |
| delta o | Paleoclimatology | 220 | 46.4 | **46.4** | $CH_4$ | Climate technology | 367 | 25.9 | 27.3 |
| sea ice | Physical science foundation | 219 | 46.1 | **46.1** | utilization | Climate technology | 258 | 23.6 | **23.6** |
| Biase | Physical science foundation | 207 | 45.9 | **45.9** | streamflow | Physical science foundation | 253 | 23.3 | 48.2 |
| divergence | Climate-change ecology | 210 | 45.7 | **45.7** | permafrost | Climate-change ecology | 205 | 22.9 | 36.1 |
| **UK** | **Topic** | **P** | **%** | **Max %** | **Australia** | **Topic** | **P** | **%** | **Max %** |
| North Atlantic | Paleoclimatology | 285 | 28.4 | 38.3 | climate change adaptation | Climate policy | 306 | 21.2 | 25.5 |
| actor | Climate policy | 249 | 26.5 | **26.5** | adaptive capacity | Climate policy | 222 | 18.0 | 34.2 |
| priority | Climate policy | 329 | 25.2 | 32.8 | health | Climate policy | 430 | 17.7 | 36.1 |
| governance | Climate policy | 271 | 24.7 | 31.4 | person | Climate policy | 756 | 15.3 | 31.6 |
| carbon emission | Climate policy | 279 | 24.0 | 27.2 | industry | Climate technology | 578 | 15.2 | 20.4 |
| initiative | Climate policy | 300 | 23.7 | 27.3 | pacific | Physical science foundation | 220 | 15.0 | 52.3 |
| intervention | Climate policy | 318 | 23.3 | 29.6 | government | Climate policy | 433 | 14.8 | 21.0 |
| target | Climate policy | 446 | 22.2 | 23.3 | wheat | Climate technology | 230 | 14.8 | 21.3 |
| society | Climate policy | 501 | 22.2 | 33.1 | livelihood | Climate policy | 218 | 14.7 | 29.8 |
| ensemble | Physical science foundation | 340 | 22.1 | 38.5 | benefit | Climate policy | 807 | 14.6 | 32.2 |
| **Germany** | **Topic** | **P** | **%** | **Max %** | **Canada** | **Topic** | **P** | **%** | **Max %** |
| North Atlantic | Paleoclimatology | 285 | 21.8 | 38.3 | permafrost | Climate-change ecology | 205 | 22.4 | 36.1 |
| delta o | Paleoclimatology | 220 | 18.6 | 46.4 | sea ice | Physical science foundation | 219 | 19.2 | 46.1 |
| pollen | Paleoclimatology | 276 | 17.8 | 22.1 | predator | Climate-change ecology | 206 | 17.0 | 50.0 |
| Atlantic | Paleoclimatology | 206 | 17.5 | 38.4 | lake | Paleoclimatology | 859 | 16.0 | 29.7 |
| core | Paleoclimatology | 464 | 17.0 | 33.2 | global climate model | Physical science foundation | 329 | 13.7 | 42.3 |
| reconstruction | Paleoclimatology | 563 | 16.9 | 33.4 | streamflow | Physical science foundation | 253 | 13.4 | 48.2 |
| instrument | Climate policy | 214 | 16.8 | 29.4 | degrees n | Physical science foundation | 248 | 12.5 | 36.3 |
| concept | Climate policy | 600 | 16.3 | 26.0 | nation | Climate policy | 249 | 12.5 | 42.2 |
| delta c | Paleoclimatology | 244 | 16.0 | 38.9 | governance | Climate policy | 271 | 12.2 | 31.4 |
| proxy | Paleoclimatology | 531 | 15.6 | 35.0 | organic carbon | Climate-change ecology | 258 | 12.0 | 34.5 |
| **France** | **Topic** | **P** | **%** | **Max %** | **Spain** | **Topic** | **P** | **%** | **Max %** |
| Atlantic | Paleoclimatology | 206 | 14.6 | 38.4 | species distribution | Climate-change ecology | 288 | 10.8 | 38.2 |
| pollen | Paleoclimatology | 276 | 14.5 | 22.1 | species richness | Climate-change ecology | 275 | 10.2 | 34.2 |
| core | Paleoclimatology | 464 | 13.6 | 33.2 | eutrophication | Climate technology | 297 | 10.1 | 15.5 |
| margin | Paleoclimatology | 287 | 13.2 | 34.2 | episode | Paleoclimatology | 338 | 9.5 | 34.6 |
| amplitude | Physical science foundation | 308 | 13.0 | 33.8 | ensemble | Physical science foundation | 340 | 9.4 | 38.5 |
| ice age | Paleoclimatology | 227 | 12.8 | 34.8 | recruitment | Climate-change ecology | 226 | 9.3 | 38.5 |
| delta o | Paleoclimatology | 220 | 12.7 | 46.4 | dispersal | Climate-change ecology | 237 | 9.3 | 41.8 |
| North Atlantic | Paleoclimatology | 285 | 12.3 | 38.3 | life cycle | Climate technology | 234 | 9.0 | 19.2 |
| episode | Paleoclimatology | 338 | 12.1 | 34.6 | leafe | Climate-change ecology | 215 | 8.8 | 21.9 |
| Holocene | Paleoclimatology | 457 | 12.0 | 26.9 | last glacial maximum | Paleoclimatology | 227 | 8.8 | 36.1 |

P: Number of publications of a country in which a term occurs; %: Number of publications of a country in which a term occurs as a percentage of the global number of publications in which the terms occurs; Max %: The maximum % for all countries.

**Figure 1 Term map of climate change research**

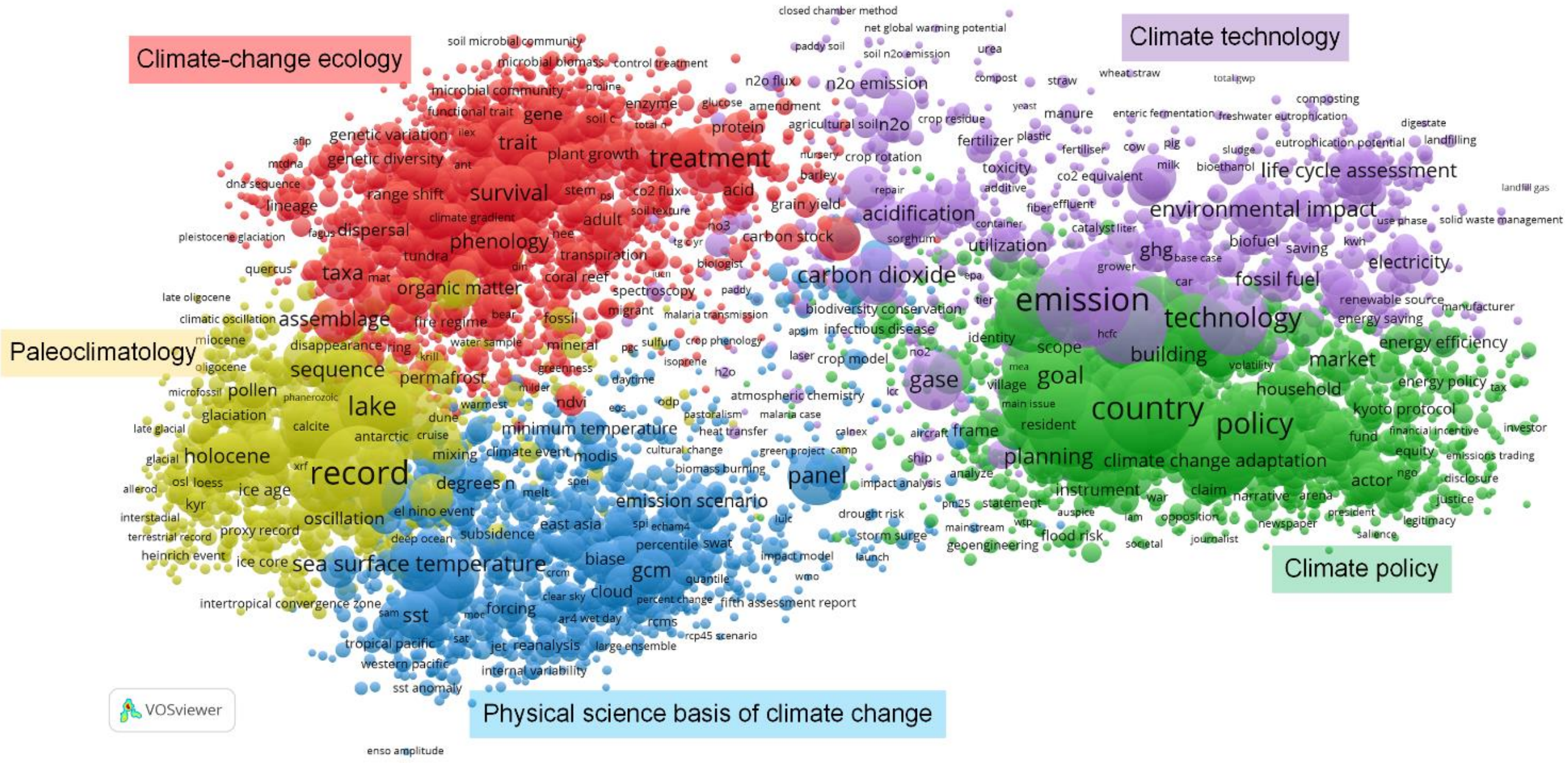

Figure 1 Map of terms extracted from the titles and abstracts of climate change articles from the period 2001-2018. Each circle represents a term. To avoid overlap of terms, some terms are not visible. Frequently occurring terms have a larger size than less frequently occurring terms. In general, the smaller the distance between two terms, the stronger their relation in terms of co-occurrences in articles. Colors represent different clusters of terms.

**Figure 2 Annual number of climate change articles between 2001 and 2018**

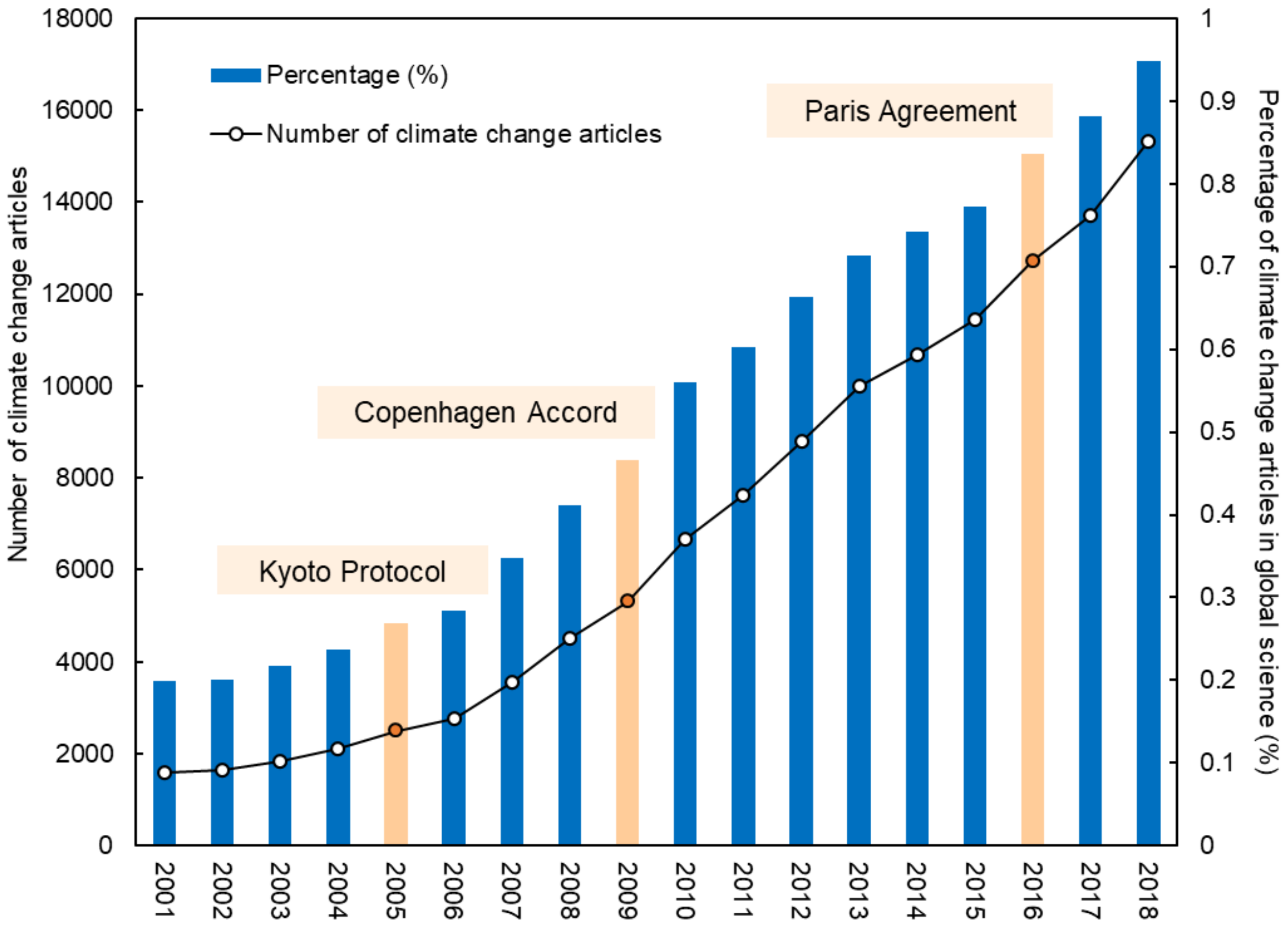

# Figure 3 Evolution of climate change research over time

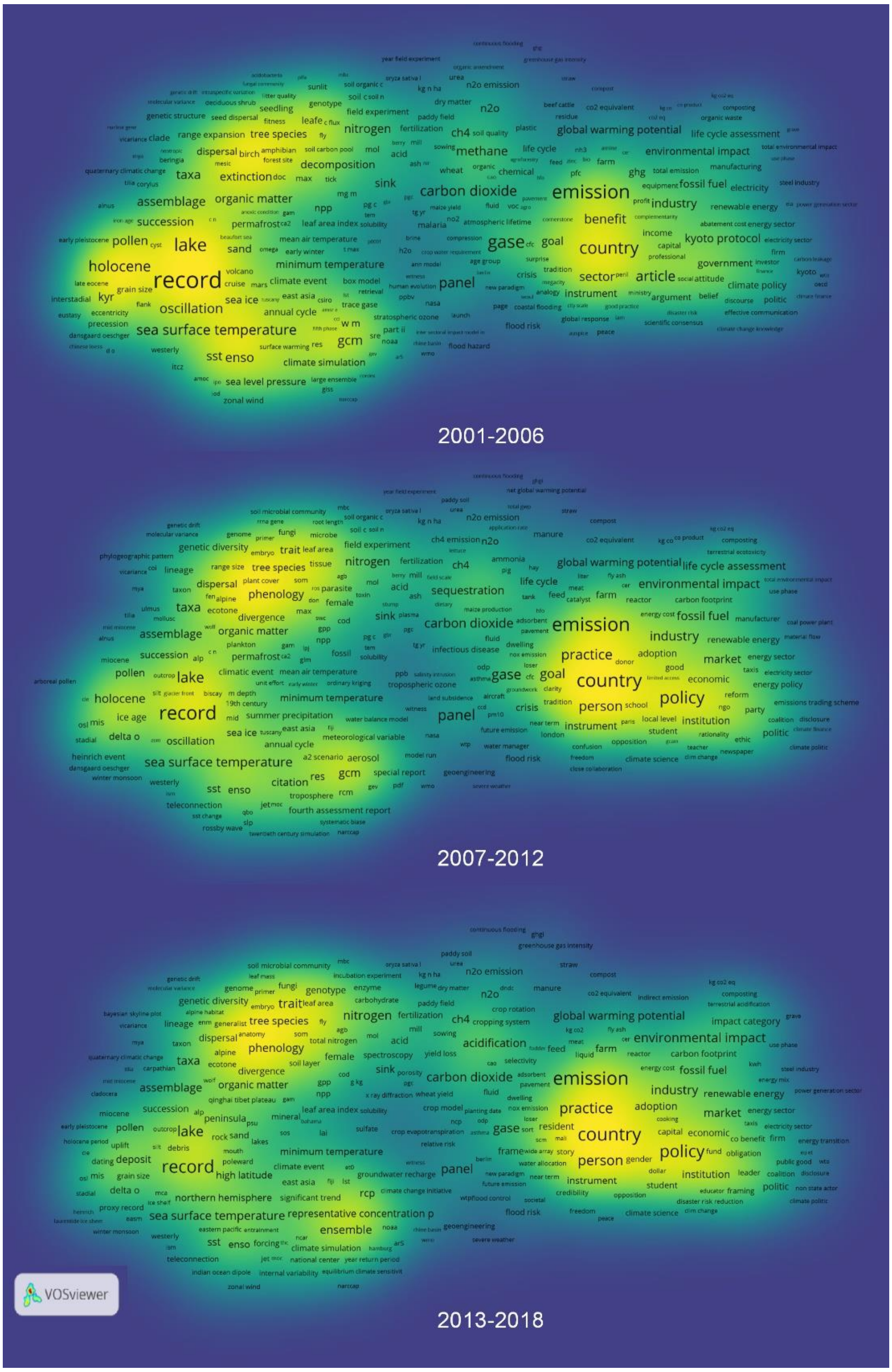

Figure 3 Density visualizations of a term co-occurrence map for three time periods. Colors indicate the focus of climate change research in a specific period, determined by the number of times each term occurs in publications from that period. The yellow regions in each visualization indicate the focal research areas in a specific period.

**Figure 4 Temporal bar graph of the top–30 author keywords burst in the period 2001–2018**

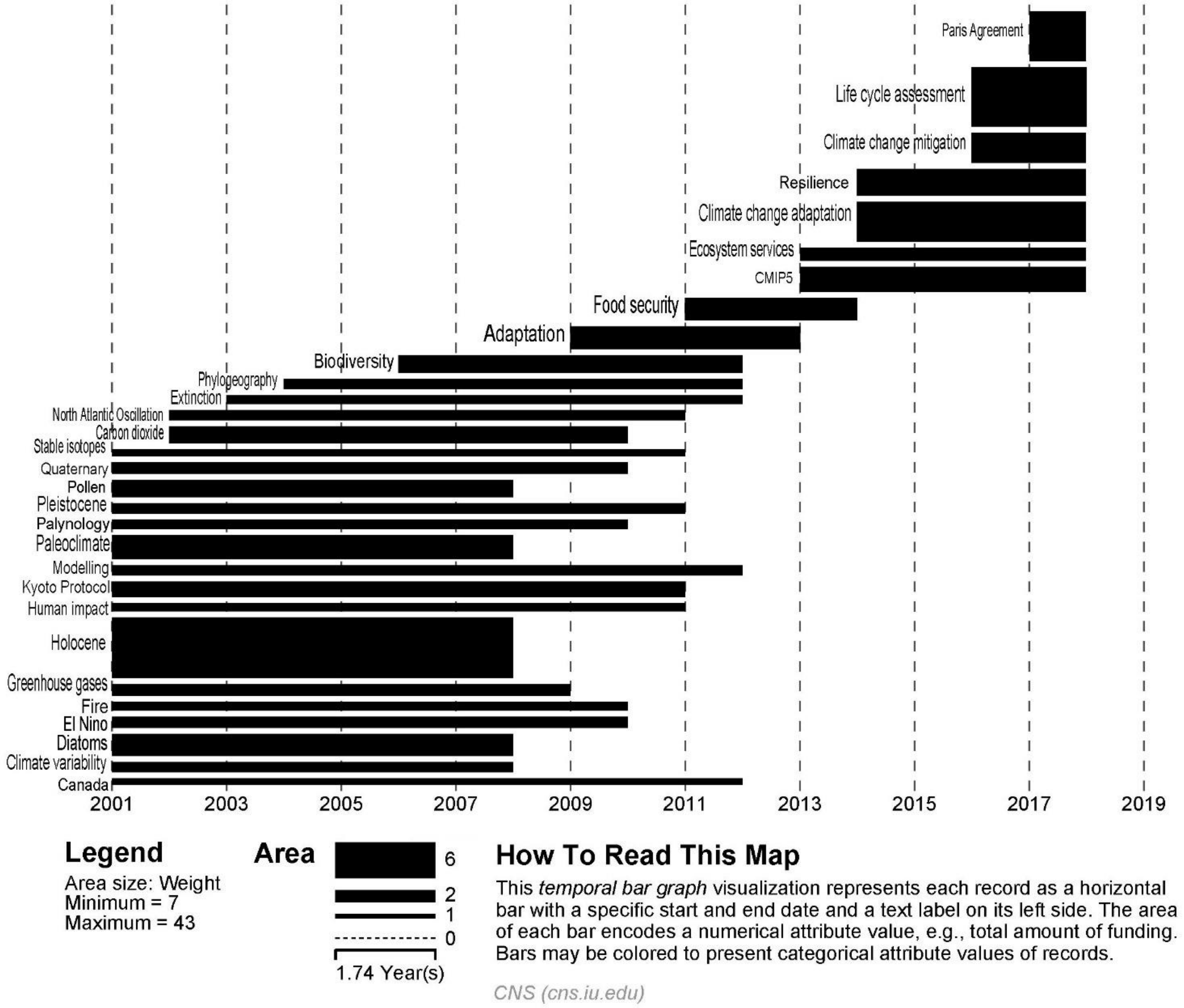

**Figure 5 Geographical distribution of climate change articles**

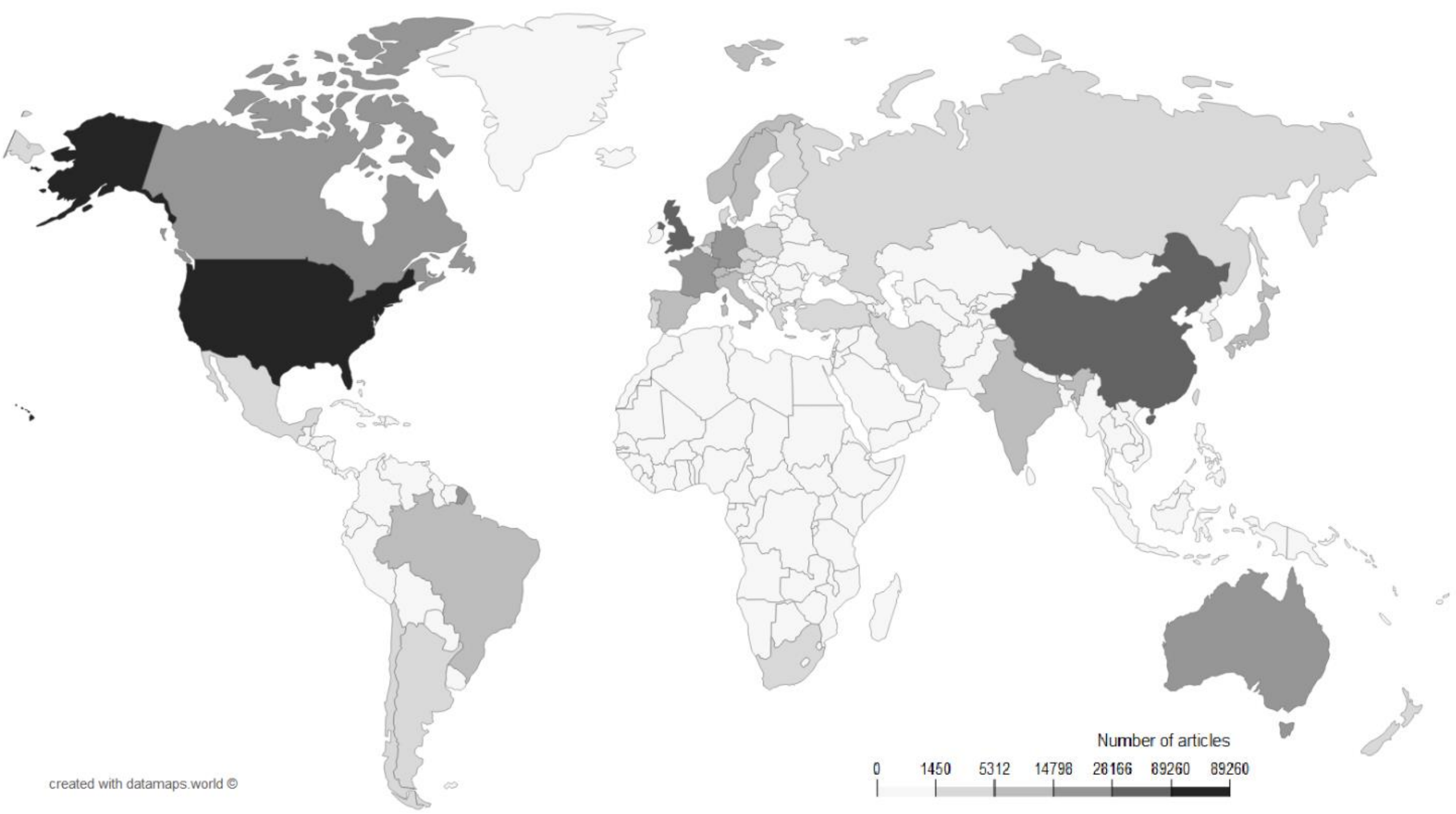

**Figure 6 Annual number of climate change articles of top eight countries between 2001 and 2018**

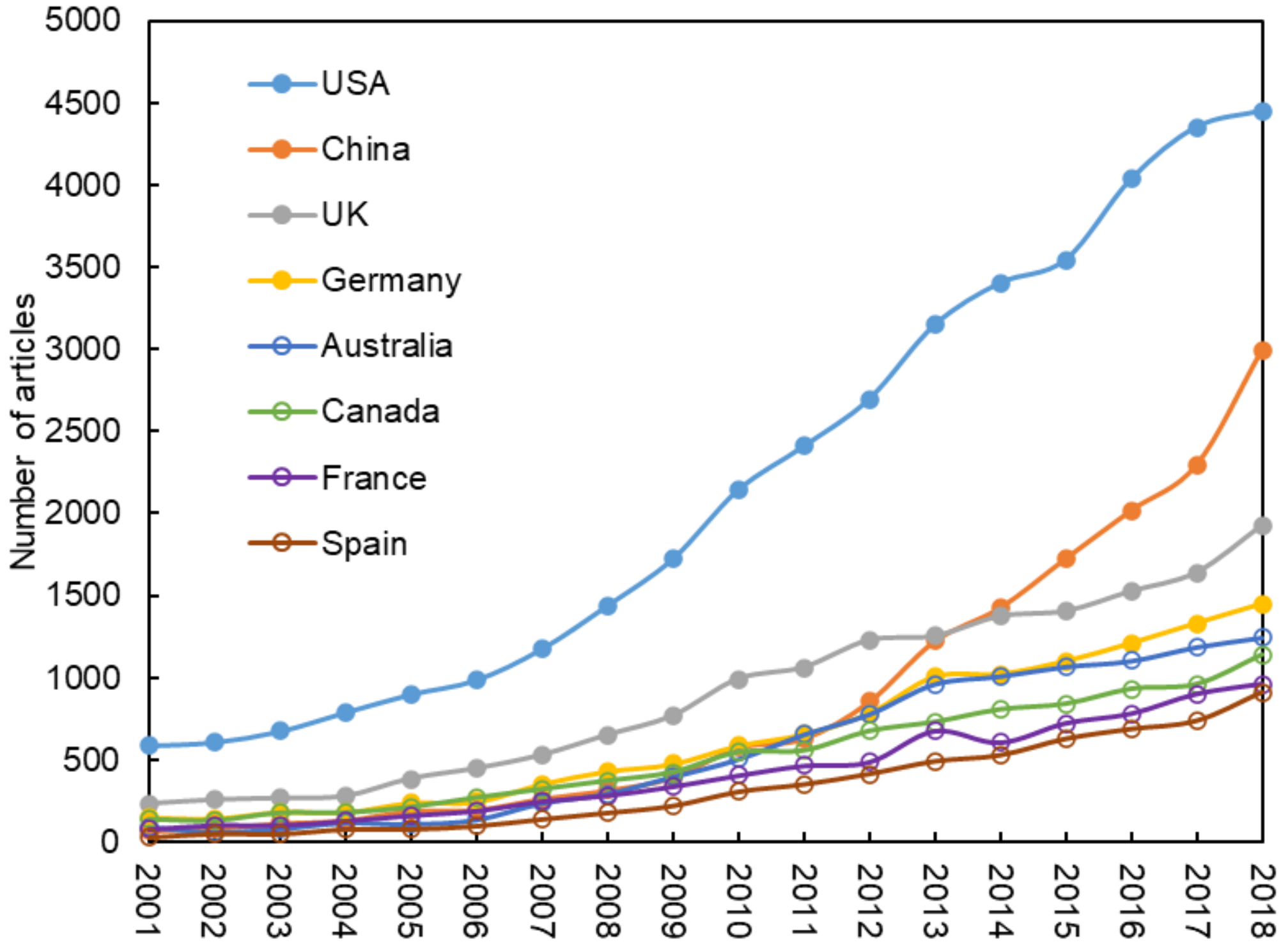

## Figure 7 Focus of climate change research of top eight countries

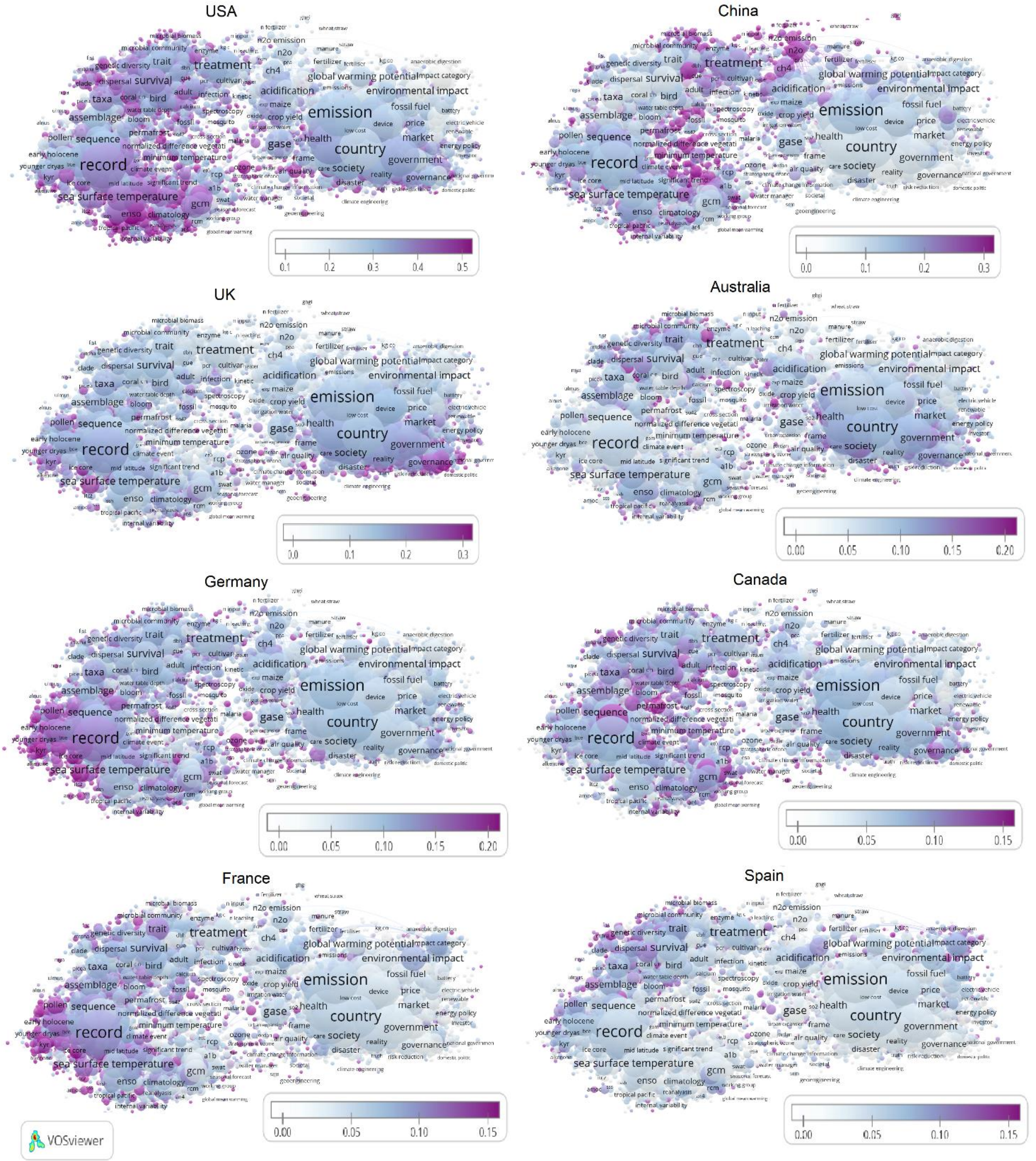

Figure 7 Overlay visualizations of a term co-occurrence map for the top eight countries. The color of a term indicates the share of the articles in which the term occurs that were produced by a specific country. Terms with a low article share are shown in white or light blue, while terms with a high article share are shown in dark purple. For example, in the case of the USA, if a term has a dark purple color, this means the USA accounts for at least half of the articles in which the term occurs.